\documentclass{pinchcr}   

\usepackage{graphicx}
\usepackage{epsfig}

\begin{document}

\title{Entanglement in many-body quantum systems}

\author{J. I. Cirac}
\affiliation{Max-Planck Institute for Quantum Optics, Hans-Kopfermannstr. 1, D-85748 Garching, Germany}

\maketitle

\section{Introduction}
\label{SectionCirac1}

Entanglement is a pure quantum property that may appear when we have a composite object. It features the existence of a very special kind of correlations, which cannot occur for product states or mixtures thereof. Those correlations, sometimes referred to as quantum correlations, give rise to a great variety of phenomena and form the basis of many applications in quantum information science \cite{NielsenChuang}. This is why, during the last fifteen years, a large theoretical effort has been devoted to define and characterize this intriguing property. More recently, some of the ideas developed in this context have been used to get a novel perspective into the many-body quantum systems that appear in Nature. In particular, the role of entanglement has been analyzed in many papers, and its implications have been used to introduce novel ways of describing such systems. In this paper we review some of the basic concepts coming from the theory of entanglement, their applications to many-body systems, and some of the new theoretical methods that have come up in this context.

Entanglement appears whenever we have two or more quantum objects. It highlights the appearance of certain kind of correlations which cannot appear in classical theories. By object we may mean a particle, a bosonic or fermionic mode, etc. Strictly speaking, the Hilbert space corresponding to the whole system must decompose as a tensor product of several Hilbert space, one for each and every object. Entanglement will depend on how we understand this decomposition and thus we must specify with respect to what (particles, modes, etc) we refer when we discuss entanglement properties of systems.
Entanglement can also be defined for both pure or mixed states. In the first case the definition of entanglement possesses no difficulty and can be very naturally understood in terms of standard correlations. In the second, however, the definition is subtle as correlations do not directly imply entanglement (in fact, correlations are everywhere in the classical world!). For two objects, entanglement can be easily defined, qualified and quantified. For many objects, however, this is not the case. New possibilities (and correlations) may appear, which cannot be reduced to the case of two objects. In fact, as of today, the theory of entanglement has not been fully developed: there exist many open questions and some definitions need to be sharpened. In the first two sections of this chapter we attempt to give a rather superficial introduction to this topic, in order to prepare the reader for the following sections. We will first consider bipartite systems (ie, two objects) and later on multipartite ones. We will introduce separately entanglement for pure and mixed states. We will highlight two quantities: the entropy of entanglement and the quantum mutual information. The first one measures the bipartite entanglement for pure states, whereas the second one measures correlations for both pure and mixed states. A thorough review of the theory of entanglement can be found in \cite{Horodecki}.

In the next section we apply the ideas previously introduced to many-body quantum systems as they appear in typical physical scenarios. In particular, we will consider spin lattices with short-range interactions and in thermal equilibrium. We will first show that certain entanglement measures display special features whenever a phase transition occurs, and explain that this is due to the simple fact that those measures basically display correlations, which are known to change abruptly under those circumstances. A thorough review of the behavior of entanglement under phase transitions can be found in \cite{Osterloh}. Then we will consider area laws in those systems. For that, we will use the quantities previously defined, namely the entropy of entanglement and the quantum mutual information. When we select a region of the lattice and consider the existing entanglement and correlations between that region and its complement one finds that they scale with the number of spins lying at the boundary of such a region. We will explain where this peculiar property comes from, and derive it for Gibbs states. There are many reviews of the area law and its application, like for example \cite{Cardy,Cramer}

In the last Section we will use the intuition developed through the area law in order to introduce an efficient way of describing many-body quantum systems. It is based on a construction in which each spin in the lattice is replaced by several auxiliary spins, which are maximally entangled to their nearest neighbors. A map is then applied to each lattice site which transforms the auxiliary spins into the original ones, giving rise to the quantum many-body state. The resulting states are known as projected entangled-pair states (PEPS), and provide such an effective description. We will highlight some of their properties, and present some explicit examples in 1D, where PEPS reduce to the so-called matrix product states (MPS), which play a central role in renormalization procedures and algorithms. Reviews on PEPS, MPS and other tensor network states can be found in \cite{FrankReview1,FrankReview2}

In this chapter, we will not make any reference to the physical system we are dealing with. This is very standard in quantum information, whereby the theory is valid for all physical quantum systems with some reachable quantum levels. Thus, many of the ideas reviewed here may well apply to atoms in optical lattices, spins in magnetic materials, or electrons in solids.

\section{Entanglement in many-body systems: Pure States}
\label{SectionCirac2}

We start out considering the simplest case, namely the entanglement present in a many-body quantum state, $\Psi$, whenever we have a pure state. This is the case, for instance, at zero temperature (if there is no degeneracy), or in most applications in quantum information science. Unfortunately, it is an idealized case since zero temperature cannot be reached in practice and, in most of the systems, interaction with the environment leads to decoherence which reflects itself in the state becoming mixed. Nevertheless, pure states are much easier to deal with than mixed states, and already display many of the features entanglement is characterized for. In this section we will concentrate in this case and consider first the situation in which we only have two systems. Later on, we will analyze the more general case of multipartite systems.

\subsection{Bipartite systems}
\label{SectionCirac21}

We consider here two systems, A and B. We denote by $H_A$ and $H_B$ the corresponding Hilbert spaces, and by $\{|n\rangle_X\}$ and orthonormal basis in $H_X$, where $n=1,2,\ldots,d_X$, with $d_X={\rm dim}(H_X)$. Most of the time we will concentrate on the simplest systems, qubits, where $d_A=d_B=2$. In that case, in order to keep the standard notation, we will take as a basis $\{|0\rangle_X,|1\rangle_X\}$. Unless we state it differently, we will always work with qubits.

The Hilbert space corresponding to the whole system, $H$, is the tensor product of $H_A$ and $H_B$, which we write $H=H_A\otimes H_B$. An orthonormal basis in that space is $\{|n\rangle_A\otimes|m\rangle_B\}$. To simplify the notation, we will typically omit the symbol "$\otimes$", and the subindices $A,B$ whenever there is no possible confusion. For instance, any state for two qubits can be written as
 \begin{equation}
 \label{Cirac_Psi}
 |\Psi\rangle = \sum_{n,m=1}^1 c_{n,m} |n,m\rangle,\quad \sum_{n,m=0}^1 |c_{n,m}|^2=1.
 \end{equation}
We will also omit the limits in the sum whenever it is obvious.

We will consider observables for each of the systems, which will be represented as operators acting on the corresponding spaces. For instance, $\sigma_1^A\otimes\sigma_2^B$ denotes an operator $\sigma_1$ acting on $A$ and $\sigma_2$ on $B$. As before, we will omit the symbol for the tensor product; additionally, $\sigma_1^A$ will stand for $\sigma_1^A\otimes 1^B$, where $1$ is the identity operator. Pauli operators acting on qubits will often appear. They are defined as follows:
 \begin{eqnarray}
 \sigma_x&=&|0\rangle\langle 1| + |1\rangle\langle 0|,\nonumber\\
 \sigma_y&=&i(|0\rangle\langle 1| - |1\rangle\langle 0|),\nonumber\\
 \sigma_z&=&|1\rangle\langle 1| - |0\rangle\langle 0|=-i\sigma_x\sigma_y,\nonumber
 \end{eqnarray}

We say that $\Psi\in H_A\otimes H_B$ is a {\em product state} if there exist two vectors $\varphi_1\in H_A$ and $\varphi_2\in H_B$ such that $|\Psi\rangle=|\varphi_1\rangle_A\otimes|\varphi_2\rangle_B$. Otherwise we say that $\Psi$ is an {\em entangled state}. Examples of product states are those forming the orhonrmal basis $|n,m\rangle$. Examples of entangled state are the so-called Bell states
 \begin{eqnarray}
 |\Phi^{\pm}\rangle &=&\frac{1}{\sqrt{2}} (|0,0\rangle\pm |1,1\rangle),\\
 |\Psi^{\pm}\rangle &=&\frac{1}{\sqrt{2}} (|0,1\rangle\pm |1,0\rangle).
 \end{eqnarray}

\subsubsection{Entanglement and correlations}
\label{SectionCirac211}

The difference between entangled and product states is that the first ones give rise to correlations. If we have a product state $|\Psi\rangle=|\varphi_1,\varphi_2\rangle$, then the expectation value $\langle \Psi|\sigma_1^A\otimes\sigma_2^B|\Psi\rangle$ factorizes into $\langle \psi_1|\sigma_1|\psi_1\rangle \langle \psi_2|\sigma_1|\psi_2\rangle$, and thus the results of measurements in both systems will be uncorrelated. For any entangled state, on the contrary, there always exist observables in $A$ and $B$ for which the expectation value does not factorize and thus for which the results of measurements will be correlated. For example, considering the Pauli operator along the direction defined by a unit vector (in the xz-plane), $\vec n$, as $\sigma_{\vec n}=n_x\sigma_x+n_z\sigma_z$, we have
 \begin{equation}
 \label{Cirac_Phi+}
 \langle\Phi^+|\sigma_{\vec n}^A\otimes \sigma_{\vec m}^B|\Phi^+\rangle={\vec n} \cdot {\vec m},
 \end{equation}
whereas $\langle\Phi^+|\sigma_{\vec n}^A|\Phi^+\rangle=0$. Thus, whenever we measure the same Pauli operators in $A$ and $B$ (ie, when $\vec n$ and $\vec m$ are parallel), the results are random but completely correlated (ie the same outcome in $A$ and $B$).

In order to highlight the power of the correlations arising in entangled states, one can consider the following game \cite{pseudotelepathy}. Two people, Alice and Bob, after meeting in order to discuss their strategy, are isolated in two rooms and given each of them a different number, $x,y=0,1$. Their goal is to output another number, $a,b=0,1$, respectively, such that $a\oplus b=xy$ (where $\oplus$ denotes addition modulo 2). That is, if $(x,y)=(0,0), (0,1)$ or $(1,0)$, they must output $a=b$, whereas for $(x,y)=(1,1)$ they must give a different output. It is very easy to show that since Alice does not know the number $y$ that Bob will receive (and viceversa), she does not know which $a$ she has to output. For instance, if Alice is given $x=1$, Bob could have $y=0$ or $y=1$, in which case she would have to give a different value of $a$. But since she does not know $y$, there is no way they can always guess. The best strategy can be easily shown to give the right answer with probability $P=0.75$. Now, if Alice and Bob during the discussion of the strategy share an entangled state $\Phi^+$, then they can guess with a probability $P=0.853... >0.75$. The strategy is: (i) for Alice to measure on her qubit the observable $\sigma_{\vec n_x}$, whenever she is given $x$, where $\vec n_0=(0,1)$, $\vec n_1=(1,0)$; (ii) for Bob, the observable $\sigma_{\vec m_y}$ whenever he is given $y$, where $\vec m_0=(1/\sqrt{2},1/\sqrt{2})$ and  $\vec m_1=(1/\sqrt{2},-1/\sqrt{2})$. It is very simple using (\ref{Cirac_Phi+}) to determine that the probability of guessing is indeed $(1+\sqrt{2})/(2\sqrt{2})$. Thus, in a world where entangled states exist, one can do things that are otherwise impossible, like to play this game with a higher probability. Actually, this is the idea behind quantum information science, where by using quantum states and correlations one can perform tasks (in the context of cryptography or computation) that are classically impossible. Furthermore, if one performs this game experimentally (for instance, using photons) one obtains a result which is incompatible with local realistic theories (ie, theories where the outcome of measurements do not depend on what is being measured somewhere else, and where the properties we measure are already well defined previous to the act of measuring). This is the essence of Bell's Theorem \cite{Bell}, which states that local realistic theories are incompatible with quantum mechanics. Note that the Bell states are the ones that gives a highest probability of wining the game, and thus they give rise to the maximal quantum correlations. They are usually referred to as maximally entangled states.

\subsubsection{Schmidt decomposition}
\label{SectionCirac212}

In order to analyze the entanglement in bipartite systems it is useful to introduce the Schmidt decomposition (SD). Given the state $\Psi$ (\ref{Cirac_Psi}), it is always possible to find an orthonormal basis $\{|u_n\rangle\}$ in $H_A$ and $\{|v_m\rangle\}$ in $H_B$, such that we can write
 \begin{equation}
 \label{Cirac_SD}
 |\Psi\rangle = \sum_{k\le d_A,d_B} d_k |u_k,v_k\rangle.
 \end{equation}
The reason is that the matrix $C$ can always be written as $C=UDV$, where $U$ and $V$ are isometries and $D$ is diagonal with positive elements. This way of writing $C$ is called singular decomposition, and it is valid for any matrix \cite{HornJohnson}. When $d_A=d_B$ the square matrices $U,V$ are unitary ($UU^\dagger=U^\dagger U=1$). Otherwise they are rectangular but still fulfill the required properties such that we can always find the Schmieedt decomposition. Note that $U$ and $V$ can be found by diagonalizing $CC^\dagger$ and $C^\dagger C$, respectively. The matrix $D$ can be found by taking the square root of the resulting diagonal matrix, which coincides in both cases. The diagonal elements of $D$ are the $d_k$ that appear in (\ref{Cirac_SD}) and are called Schmidt coefficients.

Once equipped with the SD, we can easily figure out wheater a state is entangled or not. In case one of the Schmidt coefficients is one and the rest are zero, we have a product state. Otherwise, our state is entangled. The first statement is obvious, whereas the second follows immediately when one tries to write the state as a product state.

The SD is also very useful to determine the reduced density operators for subsystems $A$ and $B$ alone. Using (\ref{Cirac_SD}) we find
 \begin{eqnarray}
 \rho_A &=&{\rm tr}_B(|\Psi\rangle\langle \Psi|)= \sum_k d_k^2 |u_k\rangle\langle u_k|\nonumber\\
 \rho_B &=&{\rm tr}_A(|\Psi\rangle\langle \Psi|)= \sum_k d_k^2 |v_k\rangle\langle v_k|\nonumber
 \end{eqnarray}
where by ${\rm tr}_X$ we mean the trace with respect to system $X$. The reduced density operators so obtained are automatically diagonalized, and we observe that their eigenvalues are nothing but the square of the Schmidt coefficients.

Let us consider the following example:
 \begin{equation}
 \label{Cirac_ExampleQubits}
 |\Psi\rangle = \cos(\theta)|0,0\rangle + \sin(\theta)|1,1\rangle,
 \end{equation}
for $\theta\in[0,\pi/4]$. This is already written in the SD form. The eigenvalues of the reduced density operator are $\cos^2(\theta)$ and $\sin^2(\theta)$. For $\theta=0$ we have a product state, whereas for $\theta=\pi/4$ we have a Bell state, which is the one that gives the largest quantum correlations, as explained in the context of the game. In parallel, the reduced density operators get more and more mixed as one increases $\theta$ from 0 to $\pi/4$ (note that the purity of a mixed state, $\rho$, is related to the distribution of its eigenvalues when considered as probabilities).

\subsubsection{Entropy of entanglement}
\label{SectionCirac213}

The previous example indicates that entanglement is related to the mixedness of the reduced density operators. In fact, this relation suggests that we can introduce a measure of entanglement by using any measure of mixedness of a state. A very natural way of the latter is the von Neumann entropy $S(\rho)=-{\rm tr}[\rho \log_2(\rho)]$ (in the context of quantum information, one defines the logarithm in base 2). Thus, we define the {\em entropy of entanglement} of a state \cite{EntropyEntanglement}, $\Psi$,
 \begin{equation}
 \label{Cirac_EntEntropy}
 E(\Psi)= -{\rm tr}[\rho_A\log_2(\rho_A)]=
 -{\rm tr}[\rho_B\log_2(\rho_B)] = -\sum_k d_k^2 \log_2(d_k^2).
 \end{equation}
As we see, this quantity can be easily determined through the SD. Note that in this expression $0\log(0)=0$ by definition. For a product state, $E=0$, whereas the maximum entanglement is $E=\log[{\min}(d_A,d_B)]$, which is reached for the state for which all the $d_k$ are equal. Those are thus called maximally entangled states (even for $d>2$).

In order to give a physical interpretation to the entropy of entanglement, we will now define two key concepts in the context of quantum information: entanglement concentration and distillation \cite{concentrationdistillation}. But, before doing that, we make a parenthesis to introduce yet another basic feature in quantum information and which is needed in order to explain the afore mentioned concepts, namely {\em generalized measurements}. The postulates of Quantum Mechanics specify what occurs when we perform a so-called filtering measurement of an observable. One just has to consider the associate operator, say $O$, and its spectral decomposition. Let us denote by $o_i$ its eigenvalues, and by $\Pi_i$ the projector onto the corresponding eigenspaces. In case $o_i$ is not degenerate, $\Pi_i=|\phi_i\rangle\langle \phi_i|$, where $O|\phi_i\rangle=o_i|\phi_i\rangle$. Note that $\sum_i \Pi=1$. Given a, in general, mixed state represented by a density operator $\rho$, the probability of obtaining the outcome $o_i$ is $P_i={\rm tr}(\Pi_i \rho)$, and the state after the measurement is $\rho_i = \Pi_i \rho \Pi_i / P_i$. Thus, a filtering measurement is characterized by a set of operators $\Pi_i$ which are positive (since they are projectors, their eigenvalues are zero or one) and add up to the identity operator. In fact, there is a more general scenario representing a measurement. We may considering bringing a measuring apparatus, letting interact with our system, and then reading off the apparatus. According to the laws of Quantum Mechanics, we will describe the measurement as follows. First, the initial state of the system and the apparatus will be $\rho\otimes|A\rangle\langle A|$, where $|A\rangle$ denotes the (pure) sate of the latter.  The interaction can be described in terms of a unitary operator, $U$, so that the state becomes $U(\rho\otimes|A\rangle\langle A|)U^\dagger$. Denoting by $p_\mu=|A_\mu\rangle\langle A_\mu|$ the projector operators defining the measurement on the apparatus, we have that the probability, $P_\mu$, of obtaining the outcome labeled by $\mu$, and the state of our system after the measurement, $\rho_\mu$, can be written as
 \begin{equation}
 \label{Cirac_GenMeas}
 p_\mu = {\rm tr}(A_\mu \rho A_\mu^\dagger),\quad \rho_\mu = \frac{A_\mu\rho A_\mu^\dagger}{p_\mu},
 \end{equation}
where $A_\mu=\langle A_\mu|U|0\rangle$ is an operator acting on our system. Note that
 \begin{equation}
 \label{Cirac_POVM}
 \sum_\mu A_\mu^\dagger A_\mu = \langle A|U\left[\sum_\mu A_\mu\rangle\langle A_\mu\right]U^\dagger |A\rangle= 1,
 \end{equation}
where we have used the fact that $|A_\mu\rangle$ is an orthonormal basis. Thus, a generalized measurement is described in terms of a set of operators, $A_\mu$, fulfilling (\ref{Cirac_POVM}). In fact, it can be easily shown that for any set of operators fulfilling that condition, there exists a measurement scheme (ie states $|A\rangle$ and $|A_\mu\rangle$) which renders (\ref{Cirac_GenMeas}) \cite{NielsenChuang}. Here, for the sake of simplicity we have restricted ourselves to pure states of the measurement apparatus and non--degenerate measurements. The extension to other cases is straightforward.

\begin{figure}[t]
\begin{center}
\includegraphics[angle=0,width=8cm]{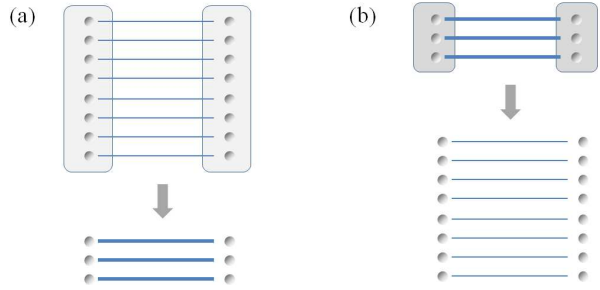}
\end{center}
\caption{(a) Entanglement distillation. Out of many copies of a weakly entangled states, by LOCC we obtain less copies of a maximally entangled state (b) Entanglement dilution. The reverse process.}
\label{Cirac_FigDistillation}
\end{figure}

Once we have defined generalized measurements, we can introduce the concept of entanglement distillation [Fig. \ref{Cirac_FigDistillation}(a)]. Let us first consider the state (\ref{Cirac_ExampleQubits}), and let us assume that our goal is to create a maximally entangled state (ie, a state in the same form but with $\theta=\pi/4$) by acting locally on each of the particles. In order to do that, we can try to apply a generalized measurement to the first particle. We choose $A_0=\tan(\theta)|0\rangle\langle 0| + |1\rangle\langle 1|$ and $A_1=(1-A_0^\dagger A_0)^{1/2}$. Those two operators fulfill (\ref{Cirac_POVM}) and this ensures that there is a physical measurement scheme associated to them which can be easily determined. In case we measure and obtain the outcome associated to $A_0$ we will achieve our goal. Otherwise, we will produce a product state instead. The probability of succeeding is $p_0=2\sin^2(\theta)$. Note that, if $A$ and $B$ are spatially separated and are hold by Alice and Bob, respectively, in order to know if the measurement has been successful, the outcome (ie a classical bit of information) has to be transmitted from Alice to Bob. One says that by local operations and classical communication (LOCC) one can distill a maximally entangled state out of the state $\Psi$ with probability $p_0$. One may wonder if there is another generalized measurement applied to A and B (individually) giving a higher probability of success. In fact, this is not the case since the generalized measurement we just chose is the optimal one. Now one can consider the case in which Alice and Bob possess two identical copies of the state $\Psi$, and they try to obtain maximally entangled states by LOCC (in which joint measurements on both qubits of Alice, or both qubits of Bob, are authorized). In general, they may get as outcome a maximally entangled state in a Hilbert space of dimension $d=2,3,4$. For instance, if they are completely successful, they will get two copies of a maximally entangled state which is equivalent to a single copy of such a state in a space of dimension $d=4$. Or if they get a single copy, they will have $d=2$. One can show that, indeed in the case of two copies the average entanglement is strictly larger than twice that for a single copy if one chooses the optimal strategy. Now we can consider what happens when we take $n$ copies and allow for the optimal LOCC in order to optimize
 \begin{equation}
 \bar E= \frac{1}{n}\sum_{d=1}^{2^n} p_d \log_2(d),
 \end{equation}
where $p_d$ is the probability they end up with an entangled state in a space of dimension $d$ (for $d=1$ they end up in a product state). The logarithm is the right quantity such that $n$ copies of a maximally entangled state (which corresponds to a dimension $2^n$) exactly gives a factor $n$. It turns out \cite{Popescudistillation} that in the limit $n\to \infty$ the result precisely coincides with the entropy of entanglement $E(\Psi)$ (\ref{Cirac_EntEntropy}). This occurs not only for qubits, but for any $d$-level systems. Thus, the entanglement entropy is nothing but the optimal averaged entanglement that we can distill out of $\Psi$ by LOCC in the asymptotic limit where we have a large number of copies.

One may consider the opposite process, called entanglement dilution [Fig. \ref{Cirac_FigDistillation}(b)]. Given $n$ maximally entangled states (of qubits), and by applying the optimal LOCC, how many copies, $m$, of the state $\Psi$ we can obtain. The ratio $\bar D= n/m$ (in average) in the limit $n\to\infty$ turns out to coincide again with $E(\Psi)$, giving again a physical meaning to the latter. In fact, in this limit, states $\Psi_{1,2}$ can be converted into each other with a yield $E(\Psi_2)/E(\Psi_1)$. This implies that, at least when we consider an scenario where we dispose of many copies of a state and are allowed to perform LOCC, there is just a single measure of entanglement, namely the entropy of entanglement.

We finish this section by mentioning other quantities that are usually employed to quantify entanglement. One is the concurrence \cite{concurrenceWooters}, which is the square of the determinant of the matrix $c$ in (\ref{Cirac_Psi}) and is nothing but the product of the Schmidt coefficients. Another one is the fidelity with a maximally entangled state, ie
 \begin{equation}
 \label{Cirac_Fidelity}
 F(\Psi) = {\rm max} |\langle \Phi^+|(U_A\otimes V_B)|\Psi\rangle|^2,
 \end{equation}
where the maximization is with respect to the unitary operators $U$ and $V$. It measures in a sense how close we are to a maximally entangled state; $U$ and $V$ just correspond to a basis change.

\subsection{Multipartite systems}
\label{SectionCirac22}

Entanglement in multipartite systems becomes more complicated than in bipartite ones. First of all, one can have that certain objects are entangled to others, but not to all of them. Second, the quantification becomes more subtle since it is not known if a property like the inter-convertibility of states by distillation and dilution exists.

We say that a state $\Psi$ of systems $A$, $B$, \ldots, $Z$, is a {\em product state} if there exists $|\varphi_X\rangle\in H_X$ such that $|\Psi\rangle=\otimes_{X} |\varphi_X\rangle_X$. Otherwise we say that we have an entangled state. Still, it may happen that some of the systems are disentangled. In order to characterize the entanglement we consider all possible partitions of the systems, and for each of them we apply the above definition. Thus, entanglement is characterized in terms of a partition which indicates which systems are entangled among themselves. For instance, for three parties we can have: (i) they are in a product state; (ii) only $A$ and $B$ are entangled; (iii) only $A$ and $C$; (iv) only $B$ and $C$; (v) all are entangled. These cases are mutually disjoint. An example of case (i) is the state $|0,0,0\rangle$, of case (ii) $|\Phi^+\rangle_{AB}\otimes|0\rangle_C$, and of case (iii) the states \cite{W,GHZ}
 \begin{eqnarray}
 \label{Cirac_WGHZ}
 |W\rangle&=&|0,0,1\rangle+|0,1,0\rangle+|1,0,0\rangle\nonumber\\
 |GHZ\rangle&=&|0,0,0\rangle + |1,1,1\rangle
 \end{eqnarray}
(we have omitted the normalization).

These two last examples illustrate the difficulty of quantifying entanglement in many-body quantum systems. It is not clear which of those states is "more" entangled. In the context of quantum information, it depends on the application we have in mind. For some, one of them is more useful, whereas for other, it is the other one. Furthermore, in this case it is not possible to convert the state $|W\rangle$ into the state $|GHZ\rangle$ by LOCC in the asymptotic limit without loosing copies (equivalently, entanglement), and thus we cannot assign a quantity like the entanglement entropy to them \cite{REGSOriginal}. One could hope to be able to convert $n$ copies of any state $\Psi$ into $m_1$ copies of $|W\rangle$ and $m_2$ of $|GHZ\rangle$, and then back in the limit $n\to\infty$, in which case one could define two measures of entanglement, given by the ratios $m_{1,2}/n$. However, this is also impossible. One may then try to include other representative states beyond those two and thus define more measures of entanglement. However, it is not even clear that a finite set of representative states exists. Thus, one cannot follow the procedure we reviewed in the previous subsection to assign a meaningful entanglement measure to multipartite states.

What we can still do is to consider bipartite partitions, in which we consider the entropy of entanglement of two disjoint sets of subsystems. This, in fact, has interesting applications in the context of many-body physics at zero temperature and will be analyzed in more detail in section \ref{SectionCirac4}.

Another approach is to look at fidelities with respect to certain particular states. For instance, one can define similarly to (\ref{Cirac_Fidelity}) the fidelity with respect to a GHZ, a W state; or to products of Bell states. Yet another possibility is to define measures which give figures of merit in specific applications. For instance, in the context of quantum repeaters, one usually defines the {\em localizable entanglement} \cite{localizable} as follows. We consider we perform measurements in all particles except for two and the goal is to obtain, in average, the maximum entanglement for those to particles. Let us consider four particles, $A,B,C,D$, and define the localizable entanglement of $\Psi$ with respect to particles $A$ and $D$. Assume we measure certain observables $O_{B,C}$ in $B$ and $C$, respectively, and denote by $|o_i\rangle$ an eigenbasis of those operators. Then, the average entanglement we obtain in $A$ and $D$ if we perform the measurement will be
 \begin{equation}
 E_{AD}(O_B,O_C) = \sum_{i,j} P_{i,j} E(\phi_{i,j}),
 \end{equation}
where $P_{i,j}$ is the probability we obtain the outcomes $i$ and $j$ in $B$ and $C$, respectively, and $\phi_{i,j}$ the state of $AD$ in such a case. This entanglement will, in general, depend on the observables we decide to measure. We thus define the localizable entanglement as the maximum with respect to all possible observables. As mentioned before, this has a specific meaning in the context of quantum repeaters whereby the goal is to obtain as much entanglement as possible between the first and the last node by measuring in the intermediate ones.

\section{Entanglement in many-body systems: Mixed States}
\label{SectionCirac3}

In this section we consider mixed states. Those are described by a density operator, $\rho$, fulfilling $\rho=\rho^\dagger\ge 0$ (meaning that all eigenvalues are non-negative), and ${\rm tr}(\rho)=1$ (normalization condition). It can always be written as
 \begin{equation}
 \label{Cirac_DO}
 \rho=\sum_i p_i |\Psi_i\rangle\langle \Psi_i|
 \end{equation}
where $p_i> 0$ fulfill $\sum_i p_i=1$ and the normalized states $\Psi_i$ need not be orthogonal. The interpretation of (\ref{Cirac_DO}) is the following: the state $\rho$ can be obtained by preparing the state $\Psi_i$ with probability $p_i$ (and then "forgetting" which state has been prepared). If one of the $p_i=1$ we have a pure state, which obviously fulfills ${\rm tr}(\rho^2)=1$; otherwise, we say that our state is mixed. Note that for mixed states there exist many ways of writing (\ref{Cirac_DO}). That is, there exist other $q_i> 0$ and $|\Phi_i\rangle$ such that $\rho$ can be decomposed in their terms. For instance,
 \begin{equation}
 \rho= \frac{1}{2}(|0\rangle\langle 0|+|1\rangle\langle 1|) = \frac{1}{2}(|+\rangle\langle +|+|-\rangle\langle -|),
 \end{equation}
where $|\pm\rangle = (|0\rangle \pm |1\rangle)/\sqrt{2}$. This simply means that the same state can be prepared in different ways, either by mixing $|0\rangle$ and $|1\rangle$, or by mixing $|\pm\rangle$.

The probability of obtaining an outcome $\mu$ when performing a (generalized) measurement is given by (\ref{Cirac_GenMeas}). Thus, the density operator contains all information we have about our system. This entails that we will not be able to distinguish by any means how we prepared a state (which decomposition we used) given that the statistics of any measurement do not depend on how we prepared the state.

In this section we will consider the entanglement of a mixed state $\rho$. We will start out with the simplest case, that of two subsystems, and later on analyze the multipartite case.

\subsection{Bipartite systems}
\label{SectionCirac32}

We consider again two subsystems, $A$ and $B$. We say that $\rho$ represents a {\em product state} whenever we can find $\rho_{A,B}$, operators acting on $H_{A,B}$ such that $\rho=\rho_A\otimes\rho_B$. A state is {\em separable} \cite{WernerSeparability} if $\rho$ can be written as a mixture of product state; that is, if
 \begin{equation}
 \label{Cirac_separable}
 \rho = \sum_i p_i |a_i,b_i\rangle\langle a_i,b_i|
 \end{equation}
where $p_i> 0$. Otherwise, we say that $\rho$ represents (or is) an {\it entangled state}.

The definition of entangled states is self-explanatory. A product state is obviously not entangled, since it does not yield correlations. A separable state cannot be entangled either, since it is a mixture of product state, which themselves are not. But separable states may contain correlations; and this is what differentiates the pure and mixed states. For example, the state
 \begin{equation}
 \label{Cirac_Examplerho}
 \rho = \frac{1}{2}(|0,0\rangle\langle 0,0|+|1,1\rangle\langle 1,1|)
 \end{equation}
fulfills $\langle \sigma_z^A\otimes \sigma_z^B\rangle=1$ whereas $\langle \sigma_z^{A,B}\rangle=0$. These correlations are, however, very trivial. If we had a classical system we could also have them. Only entangled states may display non-classical correlations.

A subtle point is that a state may look entangled even though it is separable. Let us take, for instance
 \begin{equation}
 \sigma = \frac{1}{2}(|\Phi^+\rangle\langle \Phi^+|+|\Phi^-\rangle\langle \Phi^-|)
 \end{equation}
According to this formula, we can prepare $\sigma$ by mixing two maximally entangled states. However, there is another way of preparing the same state $\sigma$ which does not require using entangled states at all. This immediately follows from the fact that $\sigma=\rho$ (one just has to replace the definition of the Bell states in this formula), and thus according to (\ref{Cirac_Examplerho}) one can prepare it by mixing two product states. The state $\sigma$ is thus separable. This simple example illustrates the difficulty of finding out whether a state is entangled or not. We have to check all possible decompositions: only if none of them involves product states we will have an entangled state. Unfortunately, there exist infinitely many decompositions, so that this task is hopeless. Fortunately, in some special cases there are shortcuts which can give us the right answer with much less effort.

\subsubsection{Entanglement witnesses}
\label{SectionCirac321}

An entanglement witness \cite{witness} is an observable which detects (witnesses) the presence of entanglement. Given an operator $W=W^\dagger$ we say that it is a witness if for all product states $|a,b\rangle$, $\langle a,b|W|a,b\rangle \ge 0$, but $W$ possesses negative eigenvalues. From the definition of separable state (\ref{Cirac_separable}), it is clear that if ${\rm tr}(\rho W)<0$, then $\rho$ must be entangled. Thus, a negative expectation value of a witness indicates the presence of entanglement. Note, however, that the converse is not necessarily true: if a the expectation value of a witness is positive, this does not imply that the corresponding state is entangled. However, one can show that for any entangled state there always exists a witness that detects it. Sadly, there is no simple way of finding out such a witness, which makes the problem of detecting entanglement in mixed state rather non-trivial.

Let us consider some examples. First, for two qubits, $W=2-S$, where $S=\sigma^A_1\otimes (\sigma^B_1+\sigma^B_2)+\sigma^A_2\otimes (\sigma^B_1-\sigma^B_2)$ and the sigma's are Pauli operators, is an entanglement witness. This can be shown by noting that for a product state, $|\langle S\rangle|\le |\langle \sigma^B_1 \rangle + \langle \sigma^B_2 \rangle| + |\langle \sigma^B_1 \rangle - \langle \sigma^B_2 \rangle|\le 2$, since $\langle \sigma\rangle \le 1$. By choosing the sigma's as in the game in Subsection \ref{SectionCirac211}, we see that for $\rho(p)= (1-p) 1/4 + p |\Phi^+\rangle\langle \Phi^+|$, ${\rm tr}[\rho(p)W]=2\sqrt{2}p$, and thus entanglement is detected for $p>1/\sqrt{2}$.

For our second example, we choose continuous variable systems. In this case, $H_{A,B}$ are infinite dimensional [and isomorphic to $L^2(R)$]. We consider two canonical operators, $X$ and $P$, for $A$ and another two, $Y$ and $Q$, for $B$. They fulfill canonical commutation relations $[X,P]=[Y,Q]=i$. We define $W=S-2$ with $S=(X-Y)^2+(P+Q)^2$. For a product state, $\langle S\rangle = \Delta X^2 + \Delta Y^2 + \Delta P^2 +\Delta Q^2 + (\langle X\rangle - \langle Y\rangle)^2 + (\langle P\rangle + \langle Q\rangle)^2\ge 2$, as a consequence of Heisenberg uncertainty relation $\Delta X^2 + \Delta P^2 \ge 1$ (and similarly for $Y$ and $Q$). On the other hand, since $[X-Y,P+Q]=0$, there always exist states for which both quantities are as small as we want, so that $W$ becomes negative. For instance, two mode squeezed states are detected by this witness.

\subsubsection{Partial transposition}
\label{SectionCirac322}

Another way of detecting entanglement is via the partial transposition \cite{perespartialtransp}. Given $\rho$, we can always write it in terms of an orthonormal basis $\{|n,m\rangle\}$, as
 \begin{equation}
 \rho = \sum_{i,j,k,l} \rho_{i,j;k,l} |i,k\rangle\langle j,l|.
 \end{equation}
We define its partial transpose with respect to $A$ in the basis $\{|n\rangle\}$,
 \begin{equation}
 \rho^{T_A} = \sum_{i,j,k,l} \rho_{i,j;k,l} |j,k\rangle\langle i,l|.
 \end{equation}
For separable states, $\rho^{T_A}$ is a valid density operator (hermitian and positive) as it can be directly shown from (\ref{Cirac_separable}). However, for entangled states this need not be the case. Thus, if $\rho^{T_A}$ has any negative eigenvalue the $\rho$ must necessarily be entangled. In particular, for two quibts it can be shown \cite{witness} that for all entangled states, the partial transposed is no longer positive semidefinite, which provides us with a very powerful tool to detect entanglement. For higher dimensional systems, however, there typically exist entangled states where $\rho^{T_A}$ is positive semidefinite. In those cases, the detection of entanglement can be very complex as one has to find an appropriate witness.

As an example, let us consider the state $\rho(p)$ defined in Subsection (\ref{SectionCirac321}). A negative eigenvalue of $\rho(p)^{T_A}$ appears whenever $p>1/3$, and thus this state is entangled in such a case and separable otherwise.

\subsubsection{Entanglement measures and mutual information}
\label{SectionCirac323}

For mixed state, entanglement measures similar to the entanglement entropy can be defined. However, they are much harder to evaluate. For instance, we can consider the distillation procedure as before but now with mixed states \cite{Vicenzodist}. That is, we may try to distill out of $n$ copies of a state $\rho$ the maximal number $m$ of states $\Phi^+$ using LOCC. The ratio $m/n$ in the limit $n\to\infty$ is called distillable entanglement, $D(\rho)$. Analogously, we may define the process of entanglement dilution and define the entanglement cost, $E_c(\rho)$. In general, $D(\rho)<E_c(\rho)$ so that we cannot first distill and then get back the same state as before. Furthermore, there exist very few examples where those quantities can be evaluated. Another measure that can be determined in practice (at least for qubits) is the entanglement of formation \cite{concurrenceWooters}
 \begin{equation}
 E_F(\rho) = {\rm min} \sum_i p_i E(\Psi),
 \end{equation}
where the minimization is done with respect to all decompositions of $\rho$ [cf (\ref{Cirac_DO})]. This quantity is related to the entanglement cost through $E_F(\rho^{\otimes n})/n\to E_c(\rho)$ in the limit $n\to\infty$.

Another way of measuring the entanglement is through the fidelity with a maximally entangled state, as in (\ref{Cirac_Fidelity}) but now with $F(\rho)={\max} \langle \Phi^+|(U\otimes V)\rho (U^\dagger\otimes V^\dagger)|\Phi^+\rangle$.

Finally, another way of measuring entanglement is using the definition of partial transposition. One defines the {\em negativity} \cite{negativity} as $N(\rho)=\max\left(||\rho^{T_A}||_1-1,0\right)$ where the 1-norm is given by the sum of the absolute values of the eigenvalues (one can also define in terms of the logarithm of such an expression). The negativity can be positive only if we have an entangled state; however there exist entangled states for which it is zero. Nevertheless, it possesses certain properties which makes it very useful to quantify entanglement in a simple way.

Another quantity of interest in the context of quantum information is the quantum mutual information, $I(A:B)$ \cite{NielsenChuang}. This does not measure entanglement, but rather correlations. In fact, it is the finest measure of correlations in the sense that it detects them even when correlation functions do not. It is defined through
 \begin{equation}
 I(A:B)= S_A + S_B -S_{AB}
 \end{equation}
Here, $S_X$ is the von Neuman entropy of $\rho$ restricted to system $X$. Thus, $S_{AB}=-{\rm tr}(\rho\log_2\rho)$, and $S_A=-{\rm tr}(\rho_A\log_2\rho_A)$, where $\rho_A={\rm tr}_B(\rho)$ is the reduced density operator of subsystem $A$. The mutual information has the following properties:
 \begin{eqnarray}
 &I(A:B)&\ge 0; \quad I(A:B)=0 \Leftrightarrow \rho=\rho_A\otimes \rho_B,\\
 \frac{1}{2}||\rho-\rho_A\otimes\rho_B||_1^2\le &I(A:B)&\le \log_2(d) ||\rho-\rho_A\otimes\rho_B||_1,\\
 &I(A:B) &\le I((aA):B) \le I(A:B)+2S_a.
 \label{Cirac_PropMutual}
 \end{eqnarray}
The first indicates that it is only zero for product states, ie when there are no correlations. The second relates it to standard correlations: whenever there are strong correlations, the mutual information is large; the converse is not entirely true, since the dimension of the Hilbert space, $d$, appears in the expression, which may be very large. Finally, the last line indicates that it decreases whenever we discard a subsystem (in this case $a$), but it cannot decrease by more than twice the entropy of such a system. The mutual information has the interpretation that is the amount of qubits we must erase in order to obtain a product state \cite{mutualinterpretation}. The proof of all those properties (except for the second one \cite{NielsenChuang}) easily follows from the strong subadditivity of the von Neumann entropy: given three subsystems, $X,Y$, and $Z$, $S_{XY}+S_{XZ} \ge S_{XYZ}+S_X$. If we choose $Y=A$, $Z=B$ and $X$ a system which is disentangled, we immediately obtain the first property. By choosing a $X=A$, $Y=B$ and $Z$ such that the whole state $\Psi$ is pure and the reduced state in $AB$ is our state $\rho$, we obtain $S_{AB}\ge S_A-S_B$, which can be used to prove the third property: taking $Z=a$ instead we obtain the first inequality, whereas using $S_{AaB}\ge S_{AB}-S_a$ and $S_a+S_A\ge S_{Aa}$ one readily obtains the second one.

\subsection{Multipartite systems}
\label{SectionCirac33}

The description of multipartite entanglement must still confront several challenges for mixed states. Whereas the definitions of product and separable states are straightforward, now we have to consider again different partitions in order to characterize this eluding property. So, given a partition of all subsystems in disjoint sets, we say that this partition is entangled if we cannot write the state as a mixture of product states along each set (but still entangled within each set; compare with multipartite pure states, and bipartite mixed states). We end up with a table in which for each partition we state whether the state is entangled or not. In order to check the entanglement for each partition, we have to find the appropriate witness, whose definition follows very naturally that for bipartite systems. Such a table contains some redundancies since, for instance, if a tripartite state is separable with respect to the partition $(A)(B)(C)$ it automatically is for any other partition. However, not all the implications which occur for pure stats concur for mixed ones. For instance, there can be tripartite states that are entangled with respect to the partitions $(AB)(C)$, $(AC)(B)$, but not for $(BC)A$ \cite{TarrachDurtripartite}. This simply means that the state may be prepared by acting together (ie letting them interact) on $A$ and $B$, and independently on $C$ (plus classical communication), or, alternatively, by acting together on $A$ and $C$, and independently on $B$. But never if $A$ is not allowed to interact with $B$ or $C$.

As measures of entanglement one can use the ones defined for mixed bipartite states properly extended to many subsystems, or those for multipartite pure states, like the localizable entanglement. However, typically one finds other measures which are more appropriate to describe specific experimental situations. In the following we briefly review the {\em spin squeezing} \cite{winelandspinsqueezing}, which plays an important role in certain precision measurements.

\subsubsection{Spin squeezing}
\label{SectionCirac331}

We consider a set of $N$ qubits, although it is simple to extend it to $d$-level systems. We denote by $\vec S=(S_x,S_y,Sz)$ the collective spin operators
 \begin{equation}
 S_\alpha = \frac{1}{2} \sum_{n=1}^N \sigma^n_\alpha.
 \end{equation}
These operators fulfill angular momentum commutation relations $[S_\alpha,S_\beta]=2i \epsilon_{\alpha,\beta,\gamma} S_\gamma$ (with $\epsilon$ the antisymmetric tensor and a sum over $\gamma$ is understood). We define the spin squeezing parameter
 \begin{equation}
 \xi = \frac{N\Delta S_z^2}{\langle S_x\rangle^2+\langle S_y\rangle^2}
 \end{equation}
It can be shown that whenever $\xi<1$, the state must be entangled \cite{spinsqueezingNature}. This quantity cannot be written in terms of an entanglement witness since it is not liner in the state. However, one can easily define in terms of a witness for two copies of the state.

For separable states, $\xi\ge 1$. This inequality is saturated when all the qubits are in the same pure state along the $XY$ plane in the Bloch sphere, for instance in $|+\rangle$. Actually, the Bloch sphere can help us to get an intuitive picture of the meaning of this quantity. There, given a state we represent the possible values of the measurement of the $S_\alpha$ by an ellipsoid centered at $\langle \vec S\rangle$, and with axis given by the corresponding variances $\Delta S_x,\Delta S_y$ and $\Delta S_z$.
For the state $|+\rangle^{\otimes N}$ we will have a circle in the $YZ$ plane centered at the point $(N/2,0,0)$ and with radius $\sqrt{N/4}$, yielding $\xi=1$. This quantity may decrease if we deform the circle into an ellipse stretched along the $Y$ direction and squeezed along $Z$. In that case, we can decrease the value of $\Delta S_z$ but keeping $\langle S_x\rangle$ practically constant (in reality it will decrease, but since it is close to $N/2\gg 1$ this will not affect much $\xi$ as long as the deformation is small). Thus, $\xi<1$ corresponds to states which are close to pure product states (ie they are close to the surface of the Bloch sphere in the $XY$ plane) but for which the variance of $S_z$ is reduced with respect to product states (at the expense of increasing some other variance). In fact, if $N$ is large and we are close to the Bloch sphere, we can use the formalism of Holstein-Primakoff \cite{HolsteinPrimakoff} in order to reexpress the situation in terms of two canonical variables. Let us assume that we work with states for which $\langle S_x\rangle\simeq N/2$; then we can treat it as a c-number (ie neglect its quantum fluctuations which will be very small as compared to its expectation value), and define two operators $X=2S_y/N$, and $P=2S_z/N$. The commutation relation $[X,P]$, when acting on the states we consider can be replaced by $[X,P]\simeq i$, and thus we are left with two canonical operators, which fulfill the uncertainty relation $\Delta X^2+ \Delta P^2\le 1$. States fulfilling $\xi<1$ are now states fulfilling $\Delta P^2<1/2$, ie the so-called squeezed states.

\subsubsection{Particle vs. mode entanglement}
\label{SectionCirac332}

We close this section by illustrating something we have already commented in the introduction; namely, that the property of entanglement depends on how we define our subsystems. Consider bosonic particles, each of them possessing two levels, $|0\rangle$ and $|1\rangle$. We can first consider the Hilbert space $H=H_A\otimes H_B\otimes\ldots$, where $H_X$ defines the space for each particle. Obviously, since we have bosons, not all states in $H$ are relevant, but just those which are symmetric under exchange of any pair of particles. On the other hand, we can consider Fork space $H_F=H_0\otimes H_1$, where $H_0$ contains states $|n\rangle_0$ with $n$ particles in state $|0\rangle$ (and similar for $H_1$). Again, if we have a fixed number of particles, not all the states will be relevant. In the first (second) case we say that we write the states in first (second) quantization. Let us consider two state written in both languages:
 \begin{eqnarray}
 |\psi_1\rangle = \frac{1}{\sqrt{2}}(|0\rangle_A\otimes|1\rangle_B+|1\rangle_A\otimes|0\rangle_B) = |1\rangle_0\otimes |1\rangle_1,\nonumber\\
 |\psi_2\rangle = \frac{1}{\sqrt{2}}(|0\rangle_A+|1\rangle_A) = \frac{1}{\sqrt{2}}(|1\rangle_0\otimes |0\rangle_1 + |0\rangle_0\otimes |1\rangle_1) ,\nonumber\\
 \end{eqnarray}
The first state is entangled in first quantization but a product in second, whereas the second one is the other way round. In fact, in the second one there is only one particle, but still we can have entanglement if we express the states in terms of modes and thus divide our space into that of the first and that of the second mode. This illustrates that, in order to properly speak about entanglement, we have to specify what are the subsystems we are considering.

\section{Entanglement and area laws}
\label{SectionCirac4}

So far we have been dealing with many-body states and their entanglement properties regardless of whether they are in equilibrium or not. We did not specify any interactions, nor temperature, but just defined some of the tools that are required to study entanglement for any state, in equilibrium or not. In this Section we concentrate on many-body states in thermal equilibrium which interact with short range interactions in lattices. This corresponds to some of the most interesting situations in experiments with cold atoms and other condensed matter systems. The requirement that we concentrate on lattices is in order to avoid some of the mathematical inconveniences of working in the continuum, although some of the ideas reviewed here may be extended to the continuum just by taking the lattice constant $a$ to zero.

Entanglement in many-body systems has been and is an active area of theoretical research in the last years. Of particular interest has been the behavior of such a property along phase transitions. We will first very briefly review some of those results, and then concentrate on a property, the area law, that seems to be fulfilled by all systems in equilibrium, as we will explain, and thus strongly characterizes many-body quantum states that appear in Nature. We will consider systems both at zero and finite temperature. In the first case, we will use the entanglement entropy and a bipartition of our system. In the latter, stronger statements can be made, where we will use the quantum mutual information.

We consider a spin system on a lattice in $d$ spatial dimensions. Some of the ideas can be extended to bosonic and fermionic systems; nevertheless, for the sake of simplicity we will use the spin language and comment on the possible extensions to fermions. Spins interact according to a Hamiltonian, $H$, which we will assume that only involves few spins (typically only two, ie we will have two-body interactions) close to each other. In other words, we will consider finite-range interactions only. We will also be interested in the thermodynamic limit where the number of spins, $N\to\infty$. We consider thermal equilibrium states at temperature $T$, ie, described by the density operator
 \begin{equation}
 \label{Cirac_Gibbs}
 \rho_T = \frac{e^{-H/T}}{Z}
 \end{equation}
where $Z={\rm tr}(e^{-H/T})$ is the partition function (and we have used units with $k_B=1$). At zero temperature, the state $\rho$ reduces to a projector onto the ground subspace, ie, that fulfilling $H|\Psi_i\rangle=E_0|\Psi_i\rangle$, where $E_0$ is the ground state energy. If the ground state is not degenerate (which we will assume in many occasions), we will just use a pure state $|\Psi_0\rangle$ to denote the ground state.

\subsection{Entanglement and phase transitions}
\label{SectionCirac41}

In thermal equilibrium, the state (\ref{Cirac_Gibbs}) will depend on external parameters driving the Hamiltonian, which we will denote with a single letter $B$, as well as on the temperature. $B$ could be an external magnetic field, interaction strengths, etc. Thus, we can write $\rho(B,T)$. As we change one of those external parameters the state may change abruptly its properties in the thermodynamic limit. Strictly speaking, some of the observables may depend non analytically on some of those parameters. At the point of non-analyticity we say that the systems undergoes a phase transition. Particularly interesting transitions may occur at $T=0$ when we change the external parameter $B$. In that case, we speak about a {\em quantum phase transition}, since quantum fluctuations are now responsible for the transition.

If we study the reduced state of two particles, $\sigma$, as we undergo a (quantum) phase transition, we may experience an abrupt behavior in its entanglement \cite{NielsenPT,FazioPT}. The reason can be easily understood as follows. Imagine we use a measure of entanglement (like the ones introduced for bipartite mixed states, Section \ref{SectionCirac323}), say $E[\sigma(B)]$. Then, if a correlation function presents a non--analyticity with respect to $B$, this will also show up as a non-analyticity in the entanglement. The reason is that $\sigma$ is an analytic function of all correlation functions, and thus whenever the latter is not analytic, then neither is $\sigma$ and thus $E$. For example (schematically),
 \begin{equation}
 \frac{dE[\sigma(B)]}{dB} =\frac{d E}{\delta \sigma} \frac{\delta\sigma}{dB},
 \end{equation}
can become discontinuous whenever $d\sigma/dB$ is, or even when the measure of entanglement is not analytic. In other words, if $E$ is not analytic at some point then it is because of the presence of a quantum phase transition, or because the definition of the entanglement measure is not analytic at that point. Thus, in some sense, looking at the entanglement of two subsystems does not really give more information about the phase transition than looking at two-body correlations. One has to look at the whole many-body system.

There exist two methods that look at the whole many-body state to study phase transitions. The first is the based on looking at quantities like $F(B_0)=\lim_{\epsilon\to 0} |\langle \Psi(B_0+\epsilon)|\Psi(B_0)\rangle|$ properly re-scaled, since they display special features whenever there is a quantum phase transition \cite{Zanardi,You}. The other method consists of using the localizable entanglement introduced in Section \ref{SectionCirac22}. There, for instance, hidden orders can be identified \cite{localizable2}.

\subsection{Area laws at zero temperature}
\label{SectionCirac42}

In this Section we will consider $T=0$ and a pure ground state of $H$. We consider a connected region $A$ of the lattice with a smooth boundary and the complementary region $B$ and concentrate on the entropy of entanglement, $E_A$, between those two regions and its growth as we make region $A$ larger and larger. According to (\ref{Cirac_EntEntropy}), this is given by the von Neumann entropy of the reduced density operator corresponding to region $A$, $\rho_A$. In general, since the entropy is an extensive quantity, one would expect that it scales with the number of spins in region $A$. In fact, this is the case if we take a random state on the lattice (according to the appropriate measure). However, for ground states of Hamiltonians as we are considering here, this seems not to be the case. Instead, the entanglement scales with the number of particles at the border of region A, $N_{\partial A}$. In a sense, the entanglement scales not with the volume of region $A$ but with the area, and thus the name area law \cite{arealaw1}.

Let us consider a simple example, the Majumdar-Ghosh \cite{MajumdarGhosh} Hamiltonian
 \begin{equation}
 \label{Cirac_MG}
 H=\sum_i \left({\vec S}^i \cdot {\vec S}^{i+1} +\frac{1}{2} {\vec S}^i \cdot {\vec S}^{i+2}\right)
 \end{equation}
where the spins are qubits and $N$ is even. Each term in the sum can be written as a projector onto the subspace of two qubis with total spin 1 plus an irrelevant constant. Thus, the ground state is a dimerized state, where singlet sates $|\Psi^-\rangle$ are formed between nearest neighbors. In fact it is doubly degenerate: one can have dimers formed between spins $2i-1$ and $2i$ or between $2i$ and $2i+1$. If we chose a set of neighboring spins as region $A$, the entropy of entanglement is bounded by 2, since at most two singlets will contribute to it. As we see, this bound does not grow with the size of $A$, and the border of $A$ has a fixed number of spins (two) which does not grow either.

\begin{figure}[t]
\begin{center}
\includegraphics[angle=0,width=5cm]{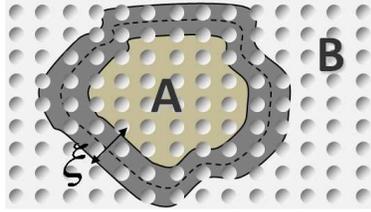}
\end{center}
\caption{Only particles that are at a distance smaller than the correlation length, $\xi$, are correlated (or entangled) to the particles in $B$. Thus, only they contribute to the entanglement entropy and therefore the area law, $E_A\le \xi N_{\partial A}$.}
\label{Cirac_FigAreaLaw}
\end{figure}

The intuition behind the area law is displayed in Fig. \ref{Cirac_FigAreaLaw}. Note that for continuous systems the entropy generally is unbounded since $N_{\delta A}\to \infty$ when we take the lattice constant $a$ to zero. Nevertheless, by scaling it properly an area law also arises.

In the last few years, the area law has been studied in a variety of systems. Here is a summary of some of the results \cite{Cardy,Cramer}: In 1-dimensional chains, it has been proven that all gapped systems fulfill the area law \cite{Hastingsarea}. For gapless systems which can be described in terms of a conformal field theory, the area law is slightly violated: $E_A\le \frac{c}{3} \log_2(\ell)$, where $\ell=L/a$ is the number of spins in region $A$, and $c$ the central charge of the conformal theory \cite{Vidalarea,CardyCFT}. Interestingly enough, this is a universal law since the coefficient in front of the logarithmic violation is a constant independent of the lattice constant. Many other models have been analyzed both analytically and numerically, and no violation (beyond the logarithmic one) has been found, at least for all reasonable models (ie homogeneous, finite range interactions, etc).

In higher dimensions, the situation is not so clear. It has been conjectured, and it seems to be corroborated by all examples studied so far, that for gapped systems the area law is fulfilled. For critical systems there is at most a logarithmic correction \cite{WolfFermionsarea,otherFermionsArea}, even though for some of them one can show that an area law is not violated \cite{Critical2Dnoviolation}. In 2-dimensions we can thus conjecture $E_A = c_1 \ell\log \ell + c_2 \ell + c_3\log^m \ell + \gamma +...$, where $\ell=L/a$ and $L$ the length of the boundary.  $c_1\ne 0$ for certain critical systems (like free fermions), and $c_3\ne 0$ for the rest. All coefficients $c_i$ are now not universal, since they change if we change the lattice constant. However, $\gamma$ can indeed be universal and, in fact, it can be connected to the existence of topological properties \cite{KitaevPreskill,LevinWen}.

As an illustration, we calculate the entanglement entropy for a set of $N$ free bosons in 1-dimension. We consider $R$ lattice sites, and a Hamiltonian for free Bosons with periodic boundary conditions
 \begin{equation}
 H= -t \sum_{r=1}^R (a^\dagger_{r} a_{r+1} + a^\dagger_{r+1} a_{r})
 \end{equation}
where $a_r$ are annihilation operators with standard commutation relations $[a_r,a^\dagger_s]=\delta_{r,s}$ and with $a_{R+1}=a_1$. The ground state is given by
 \begin{equation}
 |\Psi_0\rangle = \frac{1}{\sqrt{N!}}  \left( \frac{1}{\sqrt{R}}\sum_{r=1}^R a_r\right)^N |{\rm vac}\rangle,
 \end{equation}
with $|{\rm vac}\rangle$ the state with 0 bosons. We separate now the $R$ sites into the first $L\ll R$ and the rest, and are interested in the mode entanglement (see \ref{SectionCirac332}) between these two subsystems. We thus separate the whole Hilbert space into the tensor product of the Fock space corresponding to the first $L$ modes, and the rest. In this representation, we can write
 \begin{equation}
 |\Psi_0\rangle = \sum_{n=0}^N \sqrt{\left( \begin{array}{c} N\\ n \end{array} \right)} \left[ \frac{L}{R}\right]^{n/2} \left[\frac{R-L}{R} \right]^{(N-n)/2} |n,N-n\rangle.
 \end{equation}
From this expression it is simple to calculate the reduced density operator for the first subsystem, $\rho=\sum_{n=0}^N p(n) |n\rangle\langle n|$. For $N\gg 1$ we have
 \begin{equation}
 E_L \simeq -\int_{-\infty}^\infty p(n) \log_2[p(n)] dn,
 \end{equation}
with
 \begin{equation}
 p(n) \simeq \frac{1}{\sqrt{2\pi\sigma^2}} e^{(n-n_0)^2/2\sigma^2}
 \end{equation}
$n_0=NL/R$ and $\sigma^2=NL(R-L)/R^2$. Performing the integral, we obtain $E_A=\log_2\sigma + {\rm const}$, which in the limit $N,R\to \infty$ with $N/R$ constant it scales like $1/2\log_2 L$.

\subsection{Area laws at finite temperature}
\label{SectionCirac43}

For finite temperature, we can not longer use the entropy to derive an area law since: (i) it does not measure correlations anymore; (ii) it is an extensive quantity at any finite $T$. However, by using the quantum mutual information introduced in Section \ref{SectionCirac323} we will be able to rigorously derive \cite{HastingsMutual} an area law valid for arbitrary finite temperature $T\ne 0$ as long as we have finite-range interactions. Note that at zero temperature the mutual information reduces (up to a factor of two) to the entanglement entropy, and thus it provides us with a suitable generalization of such quantity. Nevertheless, our bound will diverge at zero temperature as it should be expected from the fact that for critical systems one obtains logarithmic corrections to the area law, as explained in the previous section.

We will consider nearest neighbor interactions, which can always be achieved under finite range interactions by redefining a spin which accumulates those of region of size equal to the interaction length. We write the Hamiltonian as
 \begin{equation}
 H= \sum_{<i,j} h_{i,j}
 \end{equation}
where $h_{i,j}$ is a Hamiltonian acting on nearest neighbor spins, and which we assume to be bounded above $||h_{i,j}||_1\le h$ by a constant. Given a region $A$ as before and its complement we want to show that
 \begin{equation}
 \label{Cirac_AreaMutual}
 I(A:B) \le N_{\delta A} \frac{h}{T}\log_2(e).
 \end{equation}
In order to show that, we will use the fact that the free energy $F(\rho)=\langle H\rangle_\rho - T S(\rho)/\log_2(e)$, where $\rho$ is any valid density operator and $S$ the von Neumann entropy, is minimized by the Gibbs state (\ref{Cirac_Gibbs}). In particular, $F(\rho_T)\le F(\rho_A\otimes\rho_B)$, where $\rho_{A,B}$ are the reduced density operators of regions $A$ and $B$, respectively. Using the definition of the free energy we obtain
 \begin{equation}
 I(A:B)=S_A+S_B-S_{AB} \le \frac{(\langle H\rangle_{\rho_T}-\langle H\rangle_{\rho_A\otimes\rho_B}}{T}\log_2(e).
 \end{equation}
By noting that the expectation value of each $h_{i,j}$ with $\rho_T$ and $\rho_A\otimes\rho_B$ coincides whenever $i$ and $j$ are both in region $A$ or $B$, we arrive to the desired expression (\ref{Cirac_AreaMutual}).

We remark that this area law is generally valid for any dimension and short-range Hamiltonian. In fact, one can also admit sufficiently fast decaying terms in the proof. Note also that for $T\to 0$ the bound diverges, at it should be according to our discussion of critical systems in the previous section.

\subsection{Area law and correlation length}
\label{SectionCirac44}

In Section \ref{SectionCirac42} we gave an intuitive picture of the area law in terms of the correlation length (see Fig. \ref{Cirac_FigAreaLaw}). In this section we will make this connection more precise using again the quantum mutual information \cite{HastingsMutual}. This result will be independent of the temperature.

\begin{figure}[t]
\begin{center}
\includegraphics[angle=0,width=9cm]{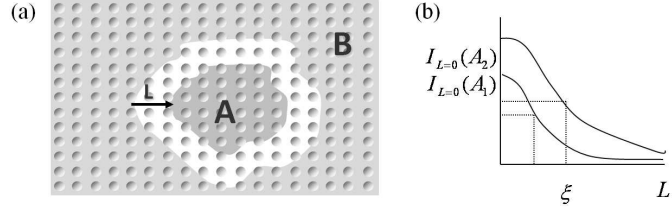}
\end{center}
\caption{(a) We choose two regions $A$ and $B$ separated by a distance $L$. (b) The mutual information decay as a function of $L$ for each choice of $A$.}
\label{Cirac_FigAreaLaw2}
\end{figure}

As we mentioned above, $I(A:B)$ measures correlations between regions $A$ and $B$. Thus, as those regions get further away, the mutual information must decay. Indeed, one of the properties of the mutual information (\ref{Cirac_PropMutual}) expressed exactly this fact. If we fix region $A$ and  separate more and more the region $B$ (see Fig. \ref{Cirac_FigAreaLaw2}), the mutual information will be reduced by a factor of two at some point, $\xi_A\in [1,\infty]$. We define the correlation length $\xi_I$ as the maximum of all $\xi_A$ with respect to all regions $A$ fulfilling the requirements we imposed above. Note that this quantity may be infinite, like for instance, in critical systems at zero temperature. Using (\ref{Cirac_PropMutual}) and the fact that in between regions $A$ and $B$ there are of the order of $N_{\partial A} L$ sites, each of them with a maximal entropy equal to $\log_2 (2s+1)$ ($s$ is the total spin), we have
 \begin{equation}
 I_{L=0}(A) \le I_{L=\xi_I}(A) + 2 N_{\partial A} \xi_I \log_2 (2s+1)
 \end{equation}
Using now that $I_{L=\xi_I}(A)\le I_{L=0}(A)/2$ we finally arrive to the area law
 \begin{equation}
 I_{L=0}(A) \le 4 N_{\partial A} \xi_I \log_2 (2s+1)
 \end{equation}
which gives a finite bond for finite correlation length.

We remark that the upper bound diverges for critical systems, as it should be given the discussion in Section \ref{SectionCirac42}. Note that the converse may not be true: if the correlation length is infinite, we may still have an area law.

\subsection{Detecting the area law}
\label{SectionCirac45}

How can we measure the area law? This seems to be difficult since measuring entropies is not an easy task, given the fact that it does not correspond to any expectation value of a physical observable. However, we may aim at measuring Renyi entropies instead, which are defined as $S_\alpha(\rho)=\log_2 ({\rm tr} \rho^\alpha)/(1-\alpha)$. In fact, for $\alpha$ integer this corresponds to measuring quantities like ${\rm tr} (\rho^n)$, which can be carried out if several copies of the system are at our disposal. The von Neumann entropy can be then determined by analytic continuation. In any case, one can use the Renyi's entropies instead of the von Neumann entropy in most of our previous definitions, and expect similar behaviors.

The quantity ${\rm tr} (\rho^2)={\rm tr}[ (\rho\otimes\rho) T^{\otimes N}]$, where $T$ is the swap operator between each site and its copy partner, ie
 \begin{equation}
 T=\sum_{n,m} |n,m\rangle\langle m,n|.
 \end{equation}
Thus, we just have to measure the observable $T$ on each particle and its copy, multiply the results, and then average over many measurements. For qubits, $T=\vec S_1\cdot \vec S_2$ up to a constant, and thus could be measured by letting the qubits interact for a while according to the Heisenberg Hamiltonian and then measure both of them. In a similar way, one can measure the $n$--th Renyi entropy by using $n$ copies. Note that one would expect that the outcome of the measurement would get exponentially small with $N$, something which should happen for random states. However, in the presence of an area law, this decay with $N$ will be softened and, in 1 spatial dimension, may allow to carry out experiments with subsystems containing large number of sites.

\section{Tensor network states}
\label{SectionCirac5}

Many-body quantum systems are very hard to describe since the number of parameters required to specify a state scales exponentially with the number of lattice sites. The reason is that the Hilbert space of the whole system is the tensor product of those corresponding to each lattice site, and thus the dimension of that space displays the exponential scaling.  In the previous section we have seen that there is a common property of thermal states in lattices systems with short range interactions, namely the area law. This fact can guide us to find out an efficient language to describe many-body quantum systems in which the number of parameters only scales polynomially with the number of sites. Under certain conditions, this is indeed possible in terms of so-called tensor network states. In this section we argue how the area law leads to such descriptions and briefly review some of them.

\subsection{Area law and Projected entangled-pair states}
\label{SectionCirac5}

\begin{figure}[t]
\begin{center}
\includegraphics[angle=0,width=9cm]{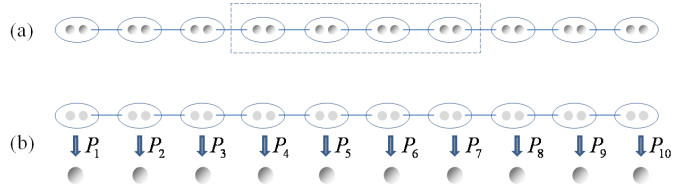}
\end{center}
\caption{(a) Dimerized state: in each node we have two spins which are maximally entangled with the neighboring spins. This is represented by a line; (b) Matrix product state construction. Starting from the dimerized state, we map the state of two spins in each site to a single spin. The maps $P$ completely characterizes the state.}
\label{Cirac_FigPEPS}
\end{figure}

Let us consider the Majumdar-Ghosh model (\ref{Cirac_MG}) which clearly illustrates how the area law arises. Let us consider a dimerized ground states, as displayed in Fig. \ref{Cirac_FigPEPS}(a). There are two spins in each site, each of them maximally entangled with the neighboring spins. The entanglement entropy of any region $A$ with the rest, $B$, equals 2, since only the entangled states at the border of the region contribute to the entanglement, each of them with one unit. If the spin Hilbert space would have dimension $D$ instead of 2, then we would have $2\log_2(D)$. Now, imagine we map each pairs of spins at each node into a single spin $s=(d-1)/2$ ($d$ is the dimension of the corresponding space), as indicated in Fig. \ref{Cirac_FigPEPS}(b). That is, we apply $P:H_D\otimes H_D\to H_d$, where $H_x$ is a Hilbert space of dimension $x$. It can be easily shown, that this map may only reduce the entanglement entropy and thus we will have $E_A\le 2\log_2(D)$, an area law. We can do exactly the same in any spatial dimension. For instance, in a square lattice, we just start with four spins at each node and use a map $P:H_D^{\otimes 4}\to H_d$. Thus, the states are generated by projecting auxiliary entangled pairs onto physical spins. The states so produced are called projected entangled-pair states (PEPS) \cite{PEPSoriginal} and obviously fulfill the area law for fixed $D$. PEPS are completely characterized by the maps $P$ we have to apply to each node, since the auxiliary states are in maximally entangled states
 \begin{equation}
 |\Phi^+\rangle=\sum_{n=1}^D |n,n\rangle.
 \end{equation}
Thus, the number of parameters required to specify a PEPS is $NdD^z$, where $z$ is the coordination number.

Let us consider a simple example. Let us consider $D=2$, one spatial dimension, and apply the map $P=|0\rangle\langle 0,0|+|1\rangle\langle 1,1|$ on each site. The resulting state is a GHZ state (\ref{Cirac_WGHZ}) of $N$ sites.

If we want to approximate in terms of PEPS the ground state of Hamiltonians with short-range interaction, as the ones considered in the previous section, we may have to increase $D$ with $N$. However, since we fulfill the area law (and may only have logarithmic violations), this increase is expected to be polynomical with $N$, rendering an efficient description of the ground states \cite{FrankEfficientMPS}. In fact, Hastings has proven \cite{HastingsPEPS} that in any spatial dimension and for any finite temperature, PEPS can be used to efficiently approximate the Gibbs state as long as the interactions are short range.

PEPS are not restricted to spin systems or pure states. In fact, one can define fermionic PEPS \cite{FermionicPEPS}, or mixed PEPS \cite{JJMixedPEPS,Zwolak}, and one can extend them even to the continuous limit \cite{cMPS} or to the case where the dimension $D\to \infty$ \cite{iMPS}. Apart from that, different algorithms have been proposed and implemented in order to solve particular Hamiltonians using those families of states variationally [see, eg, the review \cite{FrankReview1}].

\subsection{1 dimension: matrix product states}
\label{SectionCirac51}

In 1D the above construction gives rise to the so-called matrix product states (MPS) \cite{MPS1,MPS2}. By writing the map $P_n=\sum A^i_{\alpha,\beta}[n] |i\rangle\langle\alpha,\beta|$ we obtain
 \begin{equation}
 \label{Cirac_mps2}
 |\Psi\rangle_N =
 \sum_{n_1,\ldots,n_N=1}^{d}{\rm tr}\left(
 A^{n_1}[1] A^{n_2}[2]\ldots A^{n_N}[N]\right)
 \;|n_1,n_2,\ldots,n_N\rangle.
 \end{equation}
Here, $A^{n}[N]$ represents a $D\times D$ matrix. For translationally invariant systems we may choose all the matrix independent on the sites, ie $A^n[M]=A^n[M']=A^n$.

Every MPS is invariant under the exchange of $A^n_M\to X_{M-1}
A^n_M X_{M}^{-1}$, where the $X$ are non--singular matrices, as it
can be checked by direct inspection of (\ref{Cirac_mps2}). This gives us
the possibility of choosing a gauge, and thus impose conditions to
the matrices $A$ which simplify the further calculations, or which
give a physical meaning. In our case, we can impose
 \begin{equation}
 \label{Cirac_normal}
 \sum_{n=1}^{d} A^{n}[M]^\dagger A^n[M] = 1
 \end{equation}
as a gauge condition which implements such a choice.

Given a MPS, one can easily determine the expectation value of product of observables. This task reduces to calculate the trace of product of matrices, something which can be implemented very easily. In fact, one can show that all connected correlation functions decay exponentially, and the correlation length may take arbitrary values. One could expect that given the construction only near neighbor correlations exists. However, the projections may be understood as partial teleportations \cite{PEPSonewaycomputer} which can give rise to very long range correlations.

MPS lie at the realm of the density matrix renormalization group method (DMRG) \cite{DMRG}, which is widely used to determine the ground state in 1D spin chains. In fact, this method can be understood as a variational calculation with respect to the matrices $A^n[M]$. The success of this method lies on the fact that MPS can efficiently approximate the ground states of 1D chains, as explained above.

\section{Conclusions}
\label{SectionCirac6}

In this chapter, we have analyzed several entanglement properties of many-body quantum systems. In the first two sections we have reviewed the basic concepts regarding entanglement of bipartite and multipartite systems, both for pure (Section \ref{SectionCirac2}) and mixed states (Section \ref{SectionCirac3}), and related them to more standard properties (like correlations) used in quantum many-body theories. As we have argued, whereas entanglement for bipartite systems is well established, for multipartite ones there exist many possibilities of defining entanglement measures. It may turn out that one state is more entangled according to one measure, but not according to a different one. Apart from that, measures of entanglement are very hard to determine in general. Nevertheless, we have highlighted two measures of entanglement/correlations which are simple to determine and have a clear physical meaning. The first one, the entropy of entanglement, applies to bipartite pure states and is given by the von Neumann entropy of the reduced density operator of one of the subsystems. The second one, the quantum mutual information, measures the correlations between the two subsystems and for pure states it reduces to the entropy of entanglement. One can apply these two measures to multi spin systems in lattices by separating all the spins into two disjoint regions which are then considered as a bipartite system. In the particular case where we deal with ground or thermal states of short-range interaction Hamiltonians, the application of those measurements give rise to area laws, which have been the subject of Section \ref{SectionCirac4}. These laws state that the quantum mutual information between a region $A$ and its complementary one scales with the number of spins at the boundary of $A$ (and not, as one would expect, with the total number of spins in $A$). We have also mentioned that possible violations to this law appear for certain critical systems. The violation is, however, very mild and display universal properties (independent of the lattice constant, as expected for scale-invariant states). This is very peculiar, and provides us with a signature of the many-body quantum states that appear in Nature in thermal equilibrium. In fact, the area law can guide us to find efficient descriptions of thermal equilibrium states, where the number of parameters does not grow exponentially with the volume of the system. In Section \ref{SectionCirac5} we have made use of this fact and introduced certain tensor network states, the projected entangled-pair states, and explained some of its properties. In particular, in 1D they reduced to the matrix product states, that play a very important role in a numerical methods widely used in condensed matter physics.

\end{document}